\pdfoutput=1
\documentclass[twocolumn,aps,superscriptaddress,showpacs,longbibliography]{revtex4-1}
\usepackage{amsmath}
\usepackage[pdftex]{graphicx} 
\usepackage{color}
\begin{document}
\title{%
Size dependent line broadening in the emission spectra of single GaAs 
quantum dots: \\
Impact of surface charges on spectral diffusion
}
\author{Neul~Ha}
\affiliation{National Institute for Materials Science, 1-1 Namiki, Tsukuba 305-0044, Japan}
%
\author{Takaaki~Mano}
\affiliation{National Institute for Materials Science, 1-1 Namiki, Tsukuba 305-0044, Japan}
\author{Ying-Lin~Chou}
\author{Yu-Nien~Wu}
\author{Shun-Jen~Cheng}
\affiliation{Department of Electrophysics, National Chiao Tung University, Hsinchu 30050, Republic of China}
\author{Juanita~Bocquel}
\author{Paul~M.~Koenraad}
\affiliation{Eindhoven University of Technology, 5600 MB Eindhoven, The Netherlands}
%
%
\author{Akihiro~Ohtake}
\author{Yoshiki~Sakuma}
\author{Kazuaki~Sakoda}
\affiliation{National Institute for Materials Science, 1-1 Namiki, Tsukuba 305-0044, Japan}
\author{Takashi~Kuroda}
\email{kuroda.takashi@nims.go.jp}
\affiliation{National Institute for Materials Science, 1-1 Namiki, Tsukuba 305-0044, Japan}
%
\date{\today}
\begin{abstract}
Making use of droplet epitaxy, we systematically controlled the height of self-assembled GaAs quantum dots by more than one order of magnitude. 
The photoluminescence spectra of single quantum dots revealed the strong dependence of the spectral linewidth on the dot height. 
Tall dots with a height of $\sim$30~nm showed broad spectral peaks with an average width as large as $\sim5$~meV, but shallow dots with a height of $\sim$2~nm showed resolution-limited spectral lines ($\leq120$~$\mu$eV). 
The measured height dependence of the linewidths is in good agreement with Stark coefficients calculated for the experimental shape variation. 
We attribute the microscopic source of fluctuating electric fields to the random motion of surface charges at the vacuum-semiconductor interface. Our results offer guidelines for creating frequency-locked photon sources, which will serve as key devices for long-distance quantum key distribution. 
\end{abstract}
\maketitle
%
\noindent{\textbf{\textit{Introduction.}}} 
Numerous photonic applications using semiconductor quantum dots rely on the discrete and delta-function-like density of states \cite{michler_book03,*michler_book09}. However, various single dot spectroscopy studies have confirmed significant line broadening in the photoluminescence spectra that is normally much broader than the transform limited width determined by the spontaneous emission rate. The line broadening mechanism is commonly attributed to spectral diffusion, where the transition frequency randomly changes through the fluctuation of a local electric field in the vicinity of dots \cite{Empedocles_PRL96,*Empedocles_JPC99,Blanton_APL96,Seufert_APL00,Turck_PRB00}. 
The fluctuating spectral line becomes integrated into a relatively broad peak 
thanks to the long time scales of signal integration compared with those of environmental motion. 

Some progress has been made in studying the short time scale dynamics of the spectral fluctuation. Photon correlation measurement can elucidate spectrally diffusive photoluminescence with subnanosecond characteristic times \cite{Sallen_nphoto10,*Sallen_PRB11,marco_PRB12}. The correlation functions routinely show monoexponent decays, which implies efficient coupling between a single dot and a small number of environment configurations. In contrast, resonant fluorescence measurements reveal broad-band noise spectra in the $0.1$~Hz to 100~kHz range, where contributions can be expected from a large number of environment configurations \cite{Kuhlmann_NatPhys13}. Photon counting statistics of resonant fluorescence further reveal the Gaussian distribution of the random environmental shifts, which might be a consequence of the central limit theorem adopted for a large ensemble \cite{Matthiesen_SciRep14}. By contrast, high-resolution Fourier transform measurement confirms a motionally narrowed Lorentzian lineshape associated with rapid environmental fluctuation \cite{Barthlot_nphys06}. %
More recent work on field-effect devices identifies charge traps at the barrier/well interface \cite{Houel_PRL12} or impurity centers \cite{Hauck_PRB14} as a dominant field source. %
Thus, the timescales and magnitudes of spectral diffusion vary greatly depending on the sample and the measurement conditions. We still lack a global understanding of the microscopic mechanism of spectral diffusion, however it is needed for developing frequency-locked photon sources as basic elements in long-distance quantum key distribution, e.g., quantum repeaters for extending the key transmission distance. 

In this work we experimentally analyze the dependence of quantum dot morphology on environment-mediated spectral broadening. For this purpose we focus on GaAs quantum dots grown by droplet epitaxy, which enables us to continuously control the quantum dot height by more than one order of magnitude. The morphology tunability contrasts with that of traditional quantum dot growth using the Stranski-Krastanow mode, where the dot profile is essentially fixed by strain relaxation and surface energies. The spectral linewidth of a single dot emission depends strongly on the dot height. The measured height dependence agrees with that of Stark coefficients along the growth direction (normal to the sample surface). We attribute the source of the electric field fluctuation to the change in the microscopic configuration of surface charges at the vacuum-semiconductor interface \cite{Wang_APL04,Majumdar_PRB11}. Thus, morphology engineering is an alternative route to achieving narrower emitter linewidths without the need for feedback techniques to suppress spectral fluctuation \cite{Prechtel_PRX14,Hansom_APL14}. 

\begin{figure}
\includegraphics[width=7.5cm]{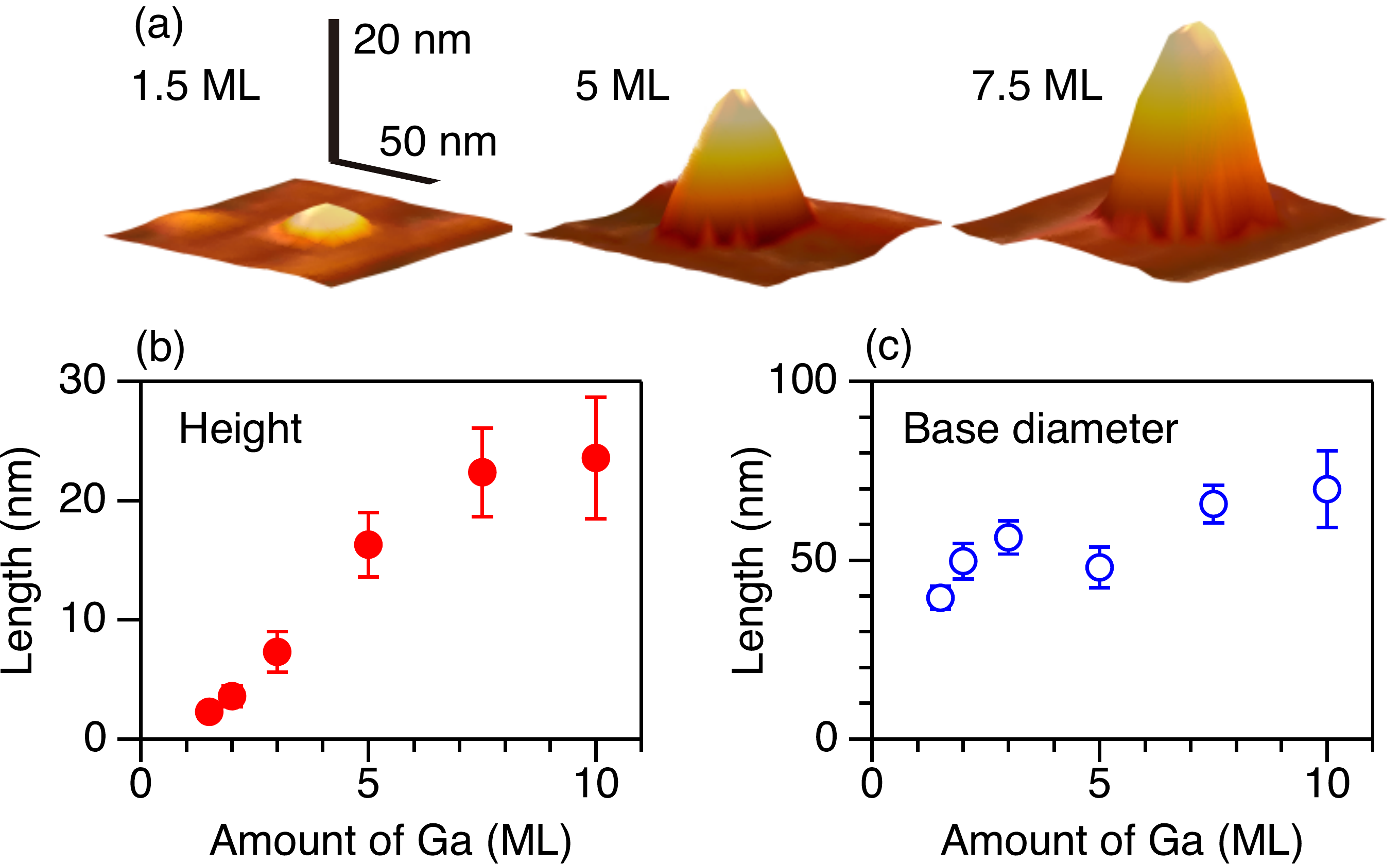}
\caption{\label{fig_afm} %
(Color online) (a) Three-dimensional AFM images for typical GaAs quantum dots grown by droplet epitaxy on Al$_{0.3}$Ga$_{0.7}$As(100) with different amounts of gallium deposition $\theta_{\mathrm{Ga}}$. (b and c) The average height and the base diameter, respectively, of quantum dots as a function of $\theta_{\mathrm{Ga}}$. The error bars represent the standard deviation of profile statistics. }
\end{figure}

\vspace{1em}
\noindent\textbf{\textit{Experimental procedure.}} 
GaAs quantum dots were self-assembly grown in Al$_{0.3}$Ga$_{0.7}$As by droplet epitaxy on semi-insulating GaAs(100) substrates \cite{Watanabe_JJAP00,mano_nanotechnology09}. These dots are free from strain thanks to the negligible lattice mismatch between GaAs and Al$_{0.3}$Ga$_{0.7}$As. 
After the growth of a 100~nm Al$_{0.3}$Ga$_{0.7}$As layer, different amounts of gallium ($\theta_{\mathrm{Ga}}$) with 1.5, 2, 3, 5, 7.5, or 10 monolayers (ML) were deposited at 0.5 ML/s and 200~$^{\circ}$C. This step enabled the formation of gallium droplets. 
Then, an As$_4$ flux was supplied at $2.5 \times 10^{-4}$ Torr and 200~$^{\circ}$C, and the gallium droplets were fully crystalized to GaAs dots. Note that the As$_4$ flux was roughly two orders of magnitude higher than that used for the self-assembly of quantum ring structures \cite{Mano_NL05,*Kuroda_PRB05}. 

After the dots were grown, the sample was annealed at 400~$^{\circ}$C \textit{in situ} (under a weak As$_4$ supply) for 10~min, and partially capped with a 20~nm Al$_{0.3}$Ga$_{0.7}$As layer. The temperature was then increased to 580~$^{\circ}$C, while the capping continued with a 30~nm Al$_{0.3}$Ga$_{0.7}$As layer followed by a 10~nm GaAs layer. 
GaAs dots on a 2~ML Al$_{0.3}$Ga$_{0.7}$As layer were additionally grown on the top of samples for atomic force microscopy (AFM) analysis. Finally, rapid thermal annealing was carried out at 800~$^{\circ}$C for 4 min in a N$_2$ atmosphere. 
All the samples with different amounts of $\theta_{\mathrm{Ga}}$ exhibited well-defined dots; see AFM top views in Supplementary Fig.~1. The dot density depended only slightly on $\theta_{\mathrm{Ga}}$ from $1.8 \times 10^{10}$~cm$^{-2}$ (1.5~ML) to $1.2 \times 10^{10}$~cm$^{-2}$ (10~ML). Thus, we assume that the volumes per dot are nearly proportional to $\theta_{\mathrm{Ga}}$. 

We used a continuous-wave laser that emitted at a wavelength of 532~nm as an excitation source. The laser illumination generated photocarriers in the Al$_{0.3}$Ga$_{0.7}$As barrier. The excitation polarization was set to be linear in order to avoid a spectral shift of nuclear origin \cite{thomas_PRB08,Sallen_NatComm14}. Our confocal setup combined an 
objective lens with a numerical aperture of $0.55$ and a hemispherical solid immersion lens (SIL) with a refractive index of two. The use of the high-index SIL enabled us to reduce the focusing diameter to $\sim0.5$~$\mu$m \cite{Yoshita_APL98}, where approximately 25 dots were inside the spot. The excitation density was kept sufficiently low so that the carrier population was less than $0.5$, \textcolor{red}{and the influence of strong optical injection on line broadening was fairly removed}. Photoluminescence signals were fed into a spectrometer of a 50~cm focusing length, and analyzed with a full width at half maximum (FWHM) resolution of 120~$\mu$eV. All the experiments were carried out at 10~K. 

\vspace{1em}
\noindent\textbf{\textit{Morphology analysis.}} 
Figure~\ref{fig_afm}(a) shows AFM three-dimensional views whose height increases significantly with the amount of $\theta_{\mathrm{Ga}}$. 
Figures~\ref{fig_afm}(b) and \ref{fig_afm}(c) show the average dot height and the base diameter, respectively, which were determined by statistical analysis. When the amount of $\theta_{\mathrm{Ga}}$ was increased from 1.5 to 10~ML, the dot height increased from $2.3 \,(\pm 0.5)$ to $24 \,(\pm 5)$ nm, i.e., by a factor of ten. 
In contrast, the base size increased only by a factor of less than two. Thus, the dot height increased considerably as the dot volume increased, while the base size remained almost unchanged. The mechanism responsible for the volume-dependent aspect ratio is explained in terms of the two-step crystallization process involved in droplet epitaxy, see Supplementary Discussion. 

\begin{figure}
\includegraphics[width=4cm]{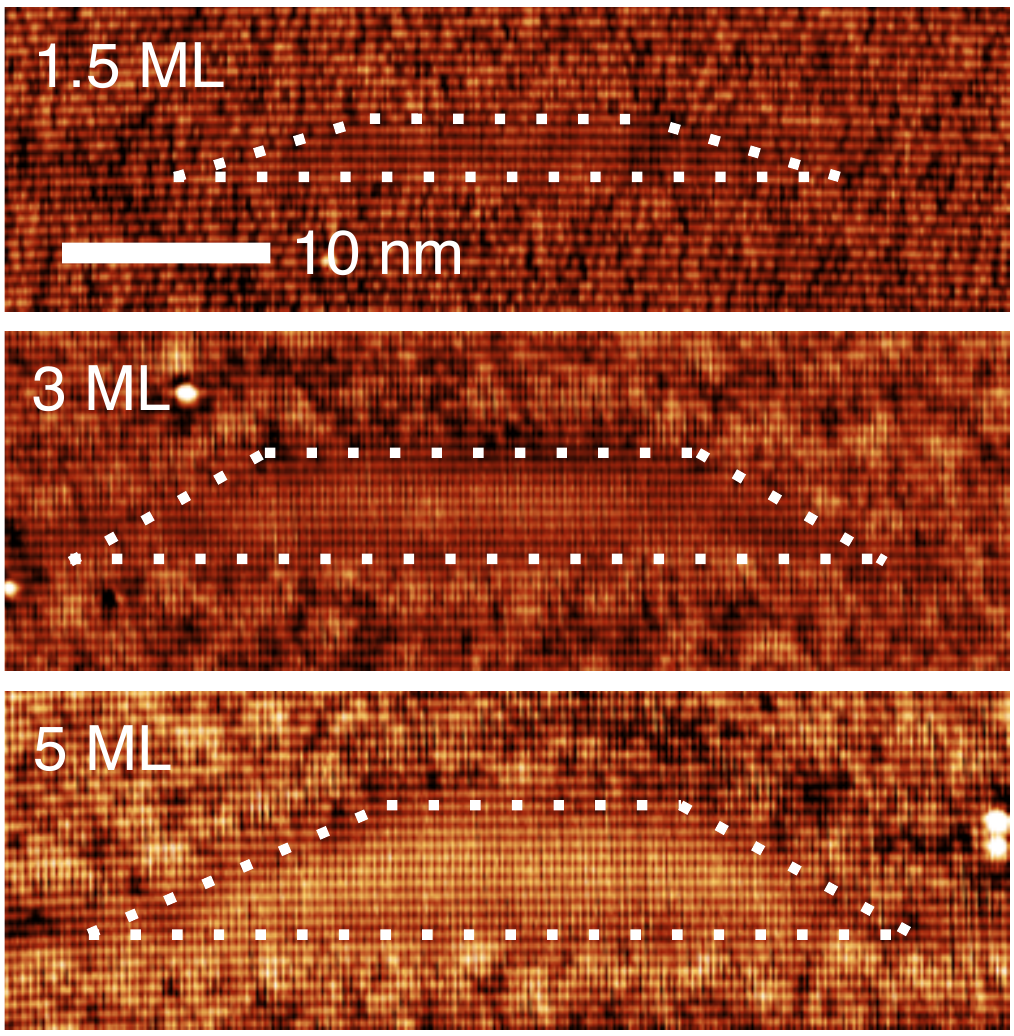}
\caption{\label{fig_STM}(Color online) Cross-sectional STM topography images of GaAs quantum dots capped with Al$_{0.3}$Ga$_{0.7}$As. The \textcolor{red}{white dotted} %
lines are guides to the eye that highlight the dot-barrier interface. }
\end{figure}

Figure~\ref{fig_STM} shows the morphology of GaAs quantum dots capped with an Al$_{0.3}$Ga$_{0.7}$As matrix that was measured using cross-sectional scanning tunneling microscopy (X-STM) \cite{Bocquel_APL14}. They have a truncated pyramidal shape, which agrees with the AFM cross-sections of uncapped dots. Thus, GaAs dots are embedded in Al$_{0.3}$Ga$_{0.7}$As while maintaining their original shape. This is due to the small diffusion length of aluminum atoms, which further gives rise to the formation of a distinct dot-barrier interface. 
This observation is in stark contrast to commonly studied InAs/GaAs dot systems, where indium atoms diffuse efficiently, and composition mixing leads to the deformation of dots with capping. 
\textcolor{red}{This shape conservation allows the determination of the shape of embedded dots from AFM measurements of free-standing references}

\begin{figure}
\includegraphics[width=6.5cm]{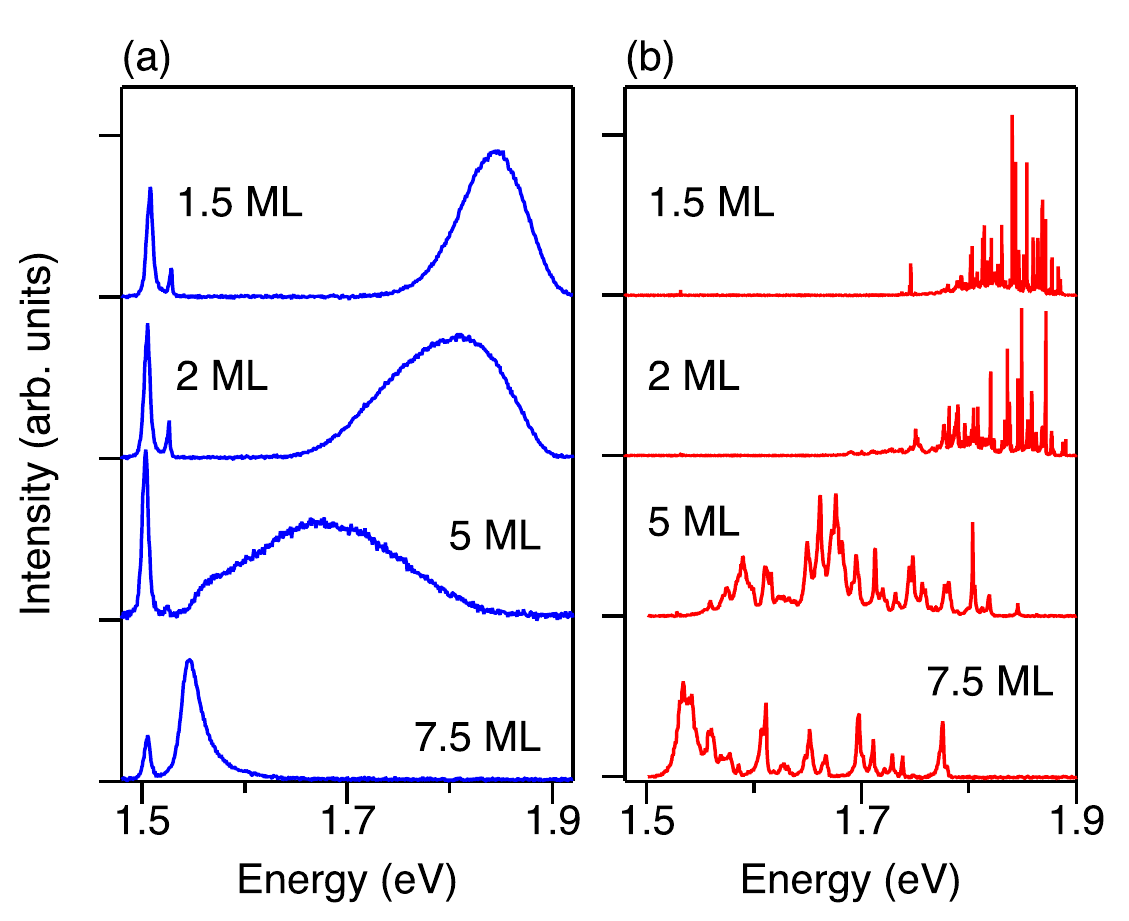}
\caption{\label{fig_PL_comparison} (Color online) Comparison of (a) photoluminescence spectra of a large ensemble of GaAs quantum dots, and (b) those of a small number of dots selected using a micro-objective setup. }
\end{figure}

\vspace{1em}\noindent\textbf{\textit{Photoluminescence spectra.}} 
Figure~\ref{fig_PL_comparison}(a) shows the spectra of a large ensemble of quantum dots. They were measured using long-focus optics. The spectral peaks at $\sim1.51$~eV originate from impurity-bound excitons in the GaAs substrate. Signals associated with quantum dots are observed at 1.85, 1.8, and 1.67~eV in the 1.5, 2, and 5~ML samples, respectively. The emission peak, therefore, shifts to a lower energy side with increasing droplet volume. The 7.5~ML sample shows a relatively narrow peak at 1.55~eV, which is close to the bulk band gap of GaAs. 

Figure~\ref{fig_PL_comparison}(b) shows the emission spectra of a small ensemble of quantum dots that were spatially selected using a micro objective setup. 
The spectra of both the 1.5 and 2 ML samples consist of sharp lines, whose linewidths are close to, or less than, the instrumental response of our spectrometer. There are around 70 spectral lines, which is approximately three times the expected number of dots inside a focusing spot. The discrepancy is reasonable because each dot is able to generate three to four emission lines through the formation of different types of charged/neutral exciton complexes \cite{marco_PRB10}. 

In contrast, the 5 and 7.5~ML samples exhibit relatively broad peaks that dominate the emission signals at energies below 1.75~eV. Note that a few sharp lines are also observed at energies higher than 1.8~eV, as found with the spectral lines of the 1.5 and 2 ML samples. Thus, the broad peaks for low-energy dots and narrow peaks for high-energy dots are not sample-specific signatures, but universal size-dependent behaviors. %

\begin{figure}
\includegraphics[width=5cm]{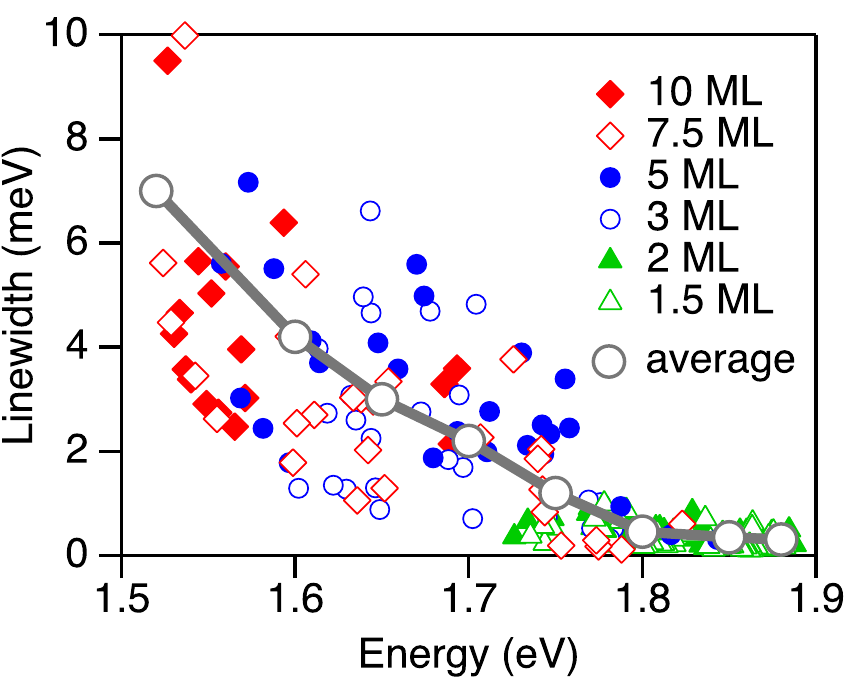}
\caption{\label{fig_width_vs_e} (Color online) Dependence of the linewidth of the spectral line on the emission energy. 
Gray open circles are the values averaged over all the spectral lines in each sample. %
}
\end{figure}

Figure~\ref{fig_width_vs_e} shows linewidth statistics as a function of emission energy. Here we evaluate the FWHM of all the spectral lines by fitting without distinguishing between the neutral and charged transitions. Such treatment is sufficient to clarify the general trend of size-dependent broadening, since the difference between the neutral and charged exciton linewidths is much smaller than the observed dot-to-dot variation, as confirmed previously \cite{marco_APL08}.
The compiled statistics demonstrate a clear correlation between line broadening and emission energy. The smooth transition over the data points of different samples confirms that the linewidth reaches several meV for tall dots, and decreases monotonically to the resolution limit with decreasing dot height. 
The similar linewidth dependence on emission energies has recently been reported for polar nitride quantum dots \cite{Kindel_PSS14}.

\vspace{1em}\noindent\textbf{\textit{Environment-induced line broadening.}} 
The effect of the environment on spectral shifts is twofold. First, hyperfine coupling between an electron and nuclei induces the Overhauser field, which acts as an effective magnetic field in the tens of mT range \cite{Urbaszek_RMP13}. Second, a charge distribution in the vicinity of dots induces a local electric field. However, the effect of nuclear fluctuation on line broadening is considered to be negligible at least in the present samples, because a typical value for a nuclear field is 10~mT, which corresponds to a spectral shift of 0.25~$\mu$eV for GaAs \cite{Sallen_NatComm14}. This is much smaller than the observed linewidth, which reaches several meV. Thus, the following discussion deals only with the effect of electric field fluctuation 
on line broadening. 

A local electric field has various microscopic origins. A common example of a field source is charge particles trapped in impurities or defects. 
However, their densities are normally very low in samples grown with molecular beam epitaxy (MBE, $\sim 10^{14}$~cm$^{-3}$). Hence, it is difficult for the bulk impurities or defects to realize line broadening that are comparable to the measured spectra, as also discussed later. %
Despite the charging and discharging of trapping centers close to dots, here we propose the fluctuation of charge carriers trapped by the vacuum-semiconductor interface as a field source. 

The formation of surface states, which trap charge carriers, is linked to the presence of 
electronically active defects at the vacuum-semiconductor interface \cite{Bardeen_PR47}. It is known that the surface state density depends on orientations and chemical treatments, and reaches $ 10^{14}$~cm$^{-2}$ for a naturally oxidized GaAs(100) surface \cite{Nannichi_JJAP88}. 
Charge carriers are efficiently trapped by the surface states, and induce a local electric field normal to the surface on average. The phenomenon also serves as the origin of band bending and Fermi-level pinning 
\cite{YuCardona}. When the sample is optically excited, some of the photoinjected carriers recombine with surface charges, and others 
occupy different surface states. Consequently, the microscopic charge arrangement changes randomly, which gives rise to field fluctuation and spectral broadening through time integration. 
The effect is orientation dependent, and taller dots become more sensitive to the induced field.

\vspace{1em}\noindent\textbf{\textit{Size-dependent Stark coefficients.}} 
Qualitative explanation of the measured line broadening is based on the derivation of Stark coefficients and the simulation of field fluctuation. The Stark shift $E_S$ of a single-particle level is described by the second-order perturbation of the interaction Hamiltonian, i.e.,  
\begin{equation}
E_S = 
\sum_{n\geq2} \frac{\left\vert \langle \psi_1 \vert eFz \vert \psi_n\rangle \right\vert^2}{E_1-E_n} 
= (eF)^2 \sum_{n\geq2} \frac{Z_{1n}^2}{E_1-E_n} \, ,
\end{equation}
where $F$ is an electric field, \textcolor{red}{$E_1$ and $E_n$ are the single particle eigen energies of the ground state and the $n$th excited state, respectively}, and $Z_{1n}=\vert \langle \psi_1 \vert z \vert \psi_n\rangle \vert$ is a dipole moment 
along a direction parallel to the field. The above equation demonstrates the size dependence of Stark shifts, where the dipole moment is proportional to the confinement length $L$, the energy denominator is scaled by $(\pi^2 h^2/2m)L^{-2}$, hence 
the Stark coefficient is enhanced as the fourth power of the effective dot size along a built-in field. 

\begin{figure}
\includegraphics[width=7cm]{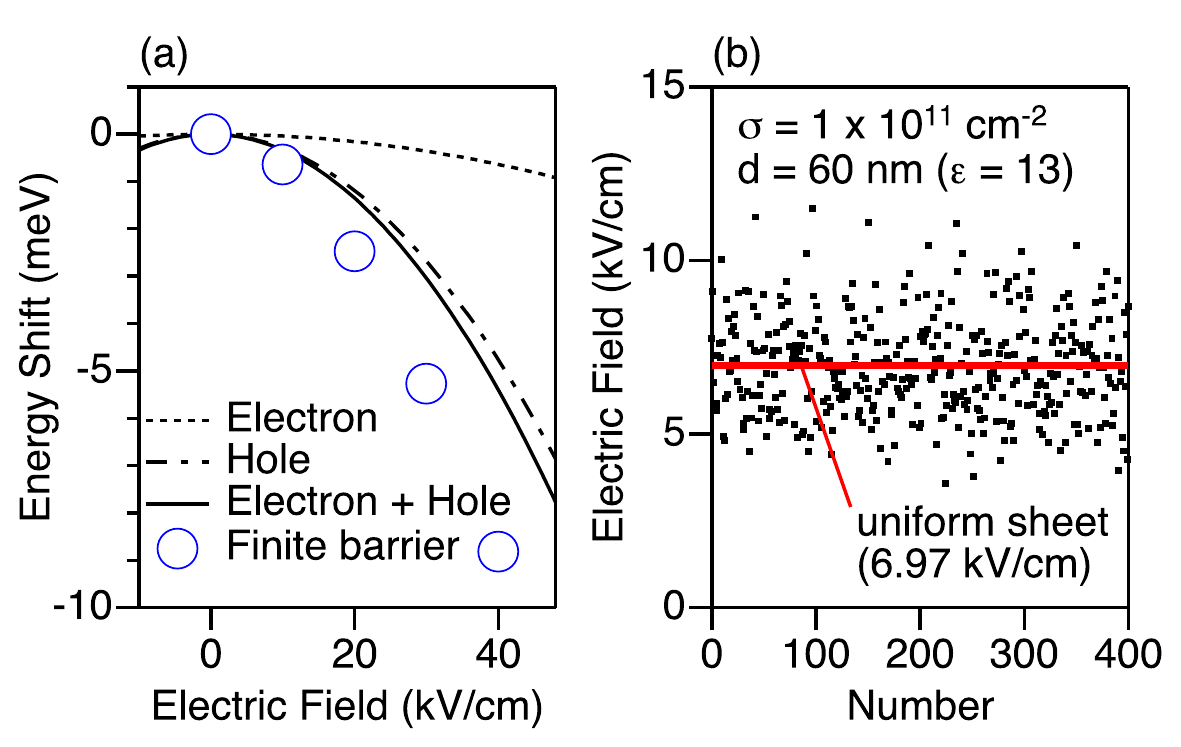}
\caption{\label{fig_Stark}(Color online) (a) Calculated Stark shift for a 12-nm-high GaAs quantum box surrounded by an infinite barrier (lines) 
and a finite barrier (open circles) as a function of a vertical field. 
(b) Monte Carlo simulation for electric fields induced by randomly positioned surface charges with a density of $1 \times 10^{11}$~cm$^{-2}$ \textcolor{red}{at a point 60 nm} from the charge layer. The red line shows a field strength induced with a uniform charge sheet. }
\end{figure}


Figure~\ref{fig_Stark}(a) shows the field dependence of spectral shifts calculated for a 12-nm-high GaAs dots. The lines show the analytic dependence for a model based on infinite-potential quantum boxes with $m^*_e \,(m^*_h) =0.067 \, (0.5)$, and the circles are the results obtained with a more precise model, which takes account of the finite GaAs/Al$_{0.3}$Ga$_{0.7}$As potential and the effect of valence-band mixing in terms of four-band $\bf{k\cdot p}$ perturbation. Both models exhibit parabolic dependence, as expected. Enhanced shifts in the finite-potential dot with respect to the infinite-potential dot arise due to the 
extended wave function. %
These results imply that an energy shift as large as 1~meV, which is a typical linewidth in the measured spectra, requires a field strength of the order of 10~kV/cm, which is expected at a position only $\sim8$~nm from a point charge. 
Accompanying impurities or defects in such close proximity to dots is fairly uncommon for MBE grown samples. This is why we have excluded bulk trapping centers and proposed surface charges as a field source.

\vspace{1em}\noindent\textbf{\textit{Simulation of field fluctuation.}} 
We evaluate the field fluctuation using a Monte Carlo simulation, where an electric field is induced by 
randomly positioned charge particles in a flat layer. Figure~\ref{fig_Stark}(b) shows the field strength distribution \textcolor{red}{at a point 60 nm} from the surface (dielectric constant $\epsilon = 13$). This condition reproduces the geometry of our structure. We found that the field changes randomly with different charge arrangements. The statistics yields a mean field strength $F_0$ of 7~kV/cm and \textcolor{red}{a standard deviation of 1.4~kV/cm} for a charge density of $1\times 10^{11}$~cm$^{-2}$. The validity of this simulation is confirmed by the agreement between the observed mean strength and the value predicted for a uniform charge sheet, $F_z = \sigma/2\pi\epsilon\epsilon_0\approx 6.97 \, \text{kV/cm}$, where $\sigma$ denotes a charge density. %
We performed the simulation for different values of $\sigma$, and found that the magnitude of field fluctuation $\Delta F$ is nearly proportional to $\sqrt{\sigma}$. 
\begin{figure}[t]
\includegraphics[width=5cm]{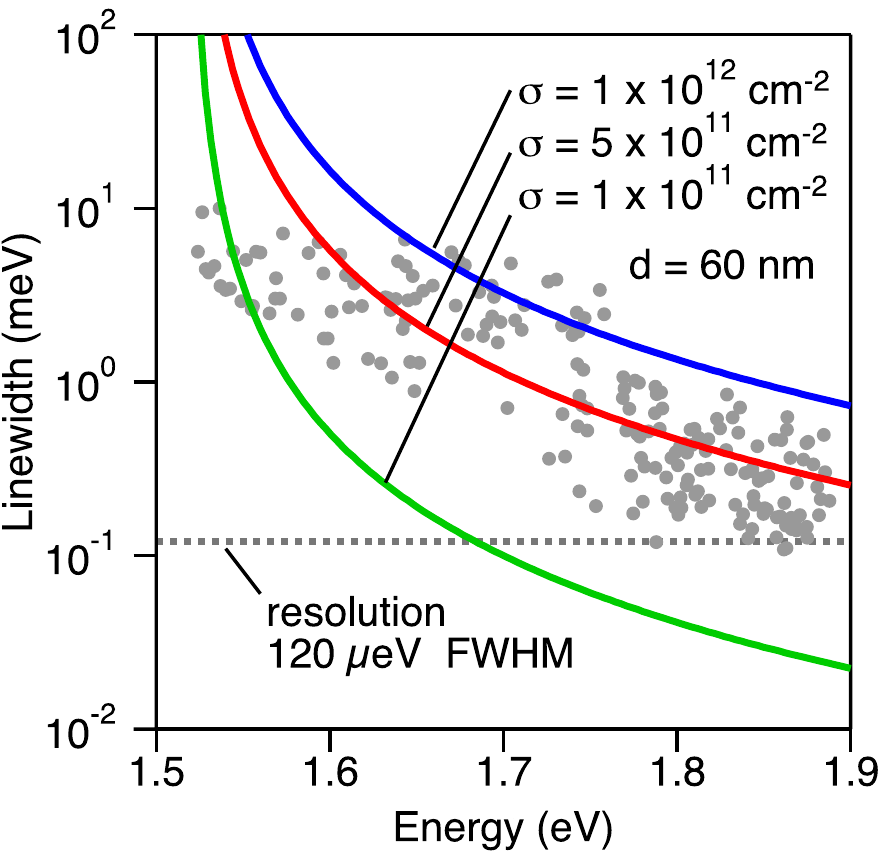}
\caption{\label{fig_th_broadening} (Color online) Calculated energy fluctuations due to randomly positioned surface charges with different densities of $1 \times 10^{11}, \, 5 \times 10^{11}$, and $1 \times 10^{12}$~cm$^{-2}$. The experimentally measured linewidths are also indicated by the gray points, \color{red}{which are equivalent to the data points shown in Fig.~\ref{fig_width_vs_e}}. }
\end{figure}

We assume that 
$\Delta F$ transfers proportionally to line broadening $\Delta E_S$, i.e., 
\begin{equation}
\Delta E_S \approx \Delta F \frac{\partial E_S(F)}{\partial F} \biggr|_{F=F_0} %
= E_S(F_0)\frac{2\Delta F}{F_0}
\, .
\end{equation}
The substitution of Eq.~1 into Eq.~2 yields the line broadening dependence on the transition energy of dots, \textcolor{red}{see Supplementary Discussion for calculation details.} Figure~\ref{fig_th_broadening} compares the experimental linewidths and the calculated spectral fluctuation for different charge densities. There is fairly good agreement between the experimental widths and calculated broadening when the charge density is of the order of $10^{11}$~cm$^{-2}$, which is a reasonable value \cite{Yablonovitch_APL87}. 

\textcolor{red}{%
Note that the lower bound of the measured linewidths is limited by our spectrometer resolution, though a previous investigation on similar dot systems using a higher-resolution spectrometer revealed the linewidths as large as a few tens of $\mu$eV at wavelengths around 1.8~eV \cite{mano_nanotechnology09}. Note also that the model curves exceed the measured linewidths at low energies. These inconsistent asymptotes are attributed to the fact that the present model ignored the effect of Coulomb binding. The quantum confined Stark shifts are associated with the wavefunction separation between electrons and holes. Coulomb attraction would inhibit such separation, and suppress energy shifts. The upper bound of line broadening is therefore roughly limited by the exciton binding energies, which were predicted to be a few tens of meV for GaAs/AlGaAs dots \cite{marco_PRB10}. %
}

\vspace{1em}\noindent\textbf{\textit{Conclusions.}} 
Spectral diffusion in the photoluminescence of single quantum dots is an interesting phenomenon that bridges microscopic random dynamics and macroscopic optical response. Here we studied morphologically controlled GaAs quantum dots grown by droplet epitaxy  to understand the source of environmental fluctuation, and demonstrated the impact of fluctuating surface charges on dot line broadening. 

From a technological point of view, however, the line broadening phenomenon is unfavorable for practical applications of quantum dots to photon emitting devices. The present results suggest several ways to engineer spectral broadening. First, we expect to suppress line broadening by creating dots with a sufficiently low aspect ratio that are robust as regards a random electric field normal to surface. 
Second, we expect to achieve narrower spectra by embedding dots more deeply in the barrier matrix, where the effect of random charges at the surface would be effectively smoothed out. Line broadening is expected to \textcolor{red}{decrease with the inverse 
of the dot-surface distance, see Supplementary Figure 3 for the simulation result}. Finally, the use of a substrate with a chemically stable surface, such as a gallium terminated (111)A surface \cite{Xu_APL09,Mano_APEX10,*Kuroda_PRB13,Ha_APL14,*Liu_PRB14}, 
and/or defect passivation technologies \cite{Nannichi_JJAP88,Yablonovitch_APL87,Robertson_JAP15} are another potential route by which to reduce surface charge fluctuation. 

This work was partially supported by Grant-in-Aid from Japan Society for the Promotion of Science. 


\bibliography{library_spectraldiffusion.bib}%
\newpage

\end{document}